\documentclass{article}
\usepackage{spconf,amsmath,graphicx,hyperref}
\usepackage{multirow,booktabs}
\usepackage{subcaption}

\title{Leveraging Multiple Speech Enhancers for Non-Intrusive Intelligibility Prediction for Hearing-Impaired Listeners}

\name{
  Boxuan Cao$^{1}$\sthanks{These authors contributed equally to this work.},
  Linkai Li$^{1,2}$\footnotemark[1], 
  Hanlin Yu$^{3}$,
  Changgeng Mo$^{1}$,
  Haoshuai Zhou$^{1}$\sthanks{Corresponding authors},
  Shan Xiang Wang$^{2}$\footnotemark[2]
  }
  
\address{
  $^1$Orka Labs Inc., China\\
  $^2$Stanford University, Electrical Engineering, United States\\
  $^3$University of British Columbia, Electrical Engineering, Canada
}
\begin{document}
\ninept
\maketitle
\begin{abstract}
Speech intelligibility evaluation for hearing-impaired (HI) listeners is essential for assessing hearing aid performance, traditionally relying on listening tests or intrusive methods like HASPI. However, these methods require clean reference signals, which are often unavailable in real-world conditions, creating a gap between lab-based and real-world assessments. To address this, we propose a non-intrusive intelligibility prediction framework that leverages speech enhancers to provide a parallel enhanced-signal pathway, enabling robust predictions without reference signals. We evaluate three state-of-the-art enhancers and demonstrate that prediction performance depends on the choice of enhancer, with ensembles of strong enhancers yielding the best results. To improve cross-dataset generalization, we introduce a 2-clips augmentation strategy that enhances listener-specific variability, boosting robustness on unseen datasets. Our approach consistently outperforms the non-intrusive baseline, CPC2 Champion across multiple datasets, highlighting the potential of enhancer-guided non-intrusive intelligibility prediction for real-world applications.

\end{abstract}
\section{Introduction}

Assessing speech intelligibility in HI listeners is central to evaluating hearing aid performance, traditionally relying on listening tests or intrusive measures such as the Hearing Aid Speech Perception Index (HASPI)\cite{HASPI}. While effective, these methods presuppose access to clean reference signals, which are rarely available in real-world conditions. This limitation underscores a critical gap between laboratory-based evaluation and real-world applicability, motivating the development of non-intrusive approaches, which offer scalable and clinically applicable intelligibility prediction without requiring reference signals \cite{2025_nonintrusive_trends}.

Recent research has explored adapting speech foundation models (SFMs) trained on large-scale corpora, such as wav2vec2.0 \cite{wav2vec2}, Whisper \cite{whisper}, HuBERT \cite{HuBERT}, and Canary \cite{canary}, for speech intelligibility prediction in hearing-impaired listeners. In the 2nd Clarity Prediction Challenge (CPC2) \cite{cpc2}, the champion model (non-intrusive) showed that combining frozen SFM backbones with prediction heads can substantially improve performance \cite{cpc2model}, highlighting a promising path toward more generalizable intelligibility prediction models. Notably, recent work suggests that best-performing prediction models rely on features from a single SFM encoder layer, rather than aggregating representations from all layers \cite{sfm_layer_selection}. 

Building on prior work adapting SFMs for non-intrusive intelligibility prediction \cite{cpc2model}, and motivated by the advantages of reference signals alongside the challenges of real-world applicability, we propose a non-intrusive speech intelligibility prediction framework that leverages strong speech enhancers to provide a parallel enhanced-signal pathway, enabling robust prediction without reference signals. We evaluated three speech enhancers and show that prediction performance strongly depends on the choice of speech enhancer, with ensembles of strong enhancers achieving the best results. Additionally, our experiments highlight significant performance degradation during cross-dataset evaluation, attributed to differences in listener profiles and recording conditions \cite{2025_nonintrusive_trends}. To improve generalization to unseen listeners and conditions, we introduce a simple 2-clips augmentation strategy that enriches listener-specific variability, leading to improved robustness on new datasets.

\section{Method}

\begin{figure*}[htb]
\begin{minipage}[b]{1.0\linewidth}
  \centering
  \raggedright{\textbf{(a)}}
  \includegraphics[width=1.0\linewidth]{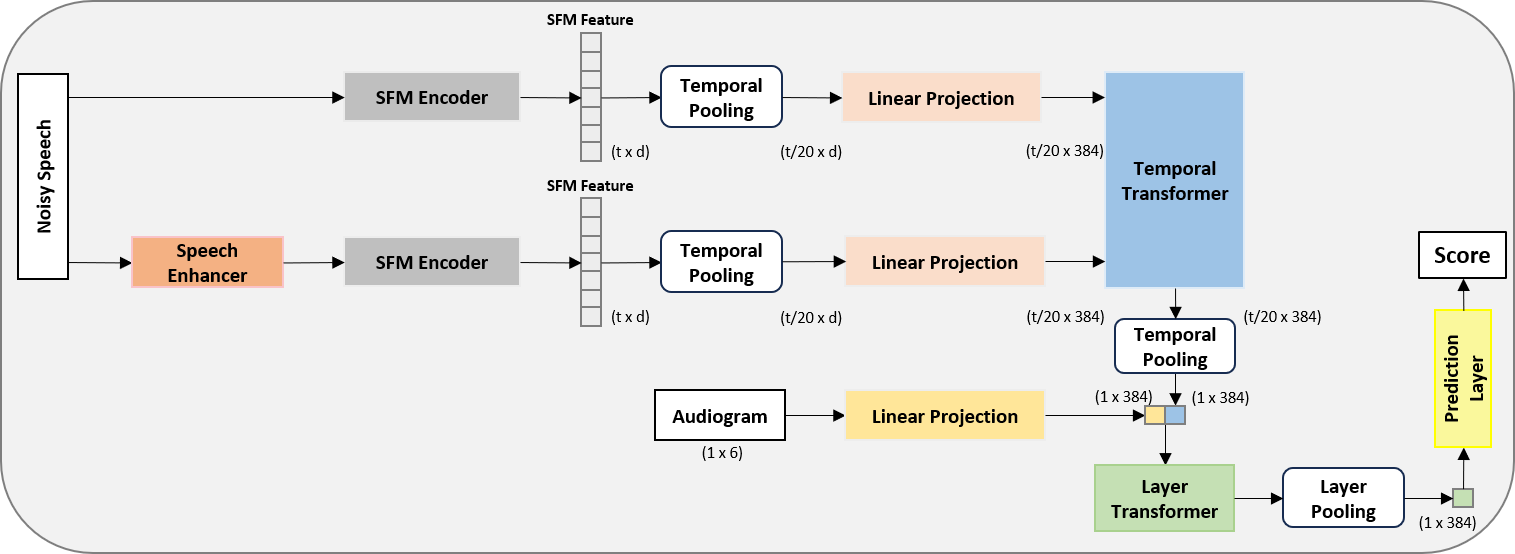}
\end{minipage}

\vspace{0.1cm} 

\begin{minipage}[b]{0.49\linewidth}
  \centering
  \raggedright{\textbf{(b)}}
  \includegraphics[width=\linewidth]{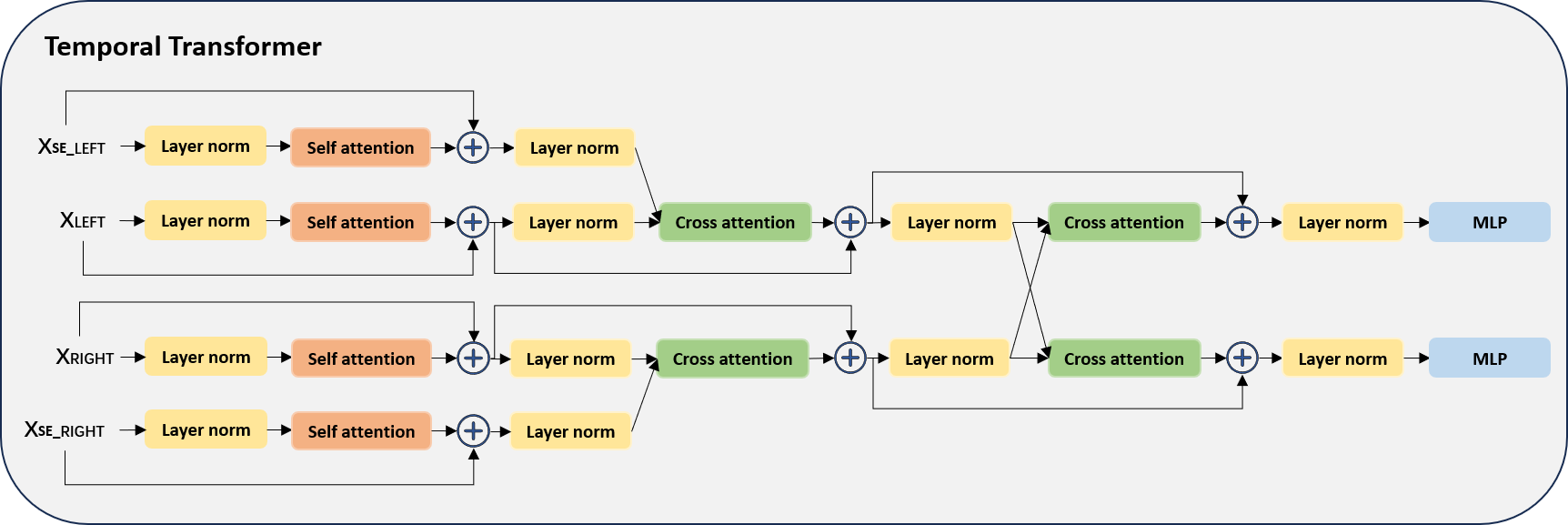}
\end{minipage}
\hfill
\begin{minipage}[b]{0.49\linewidth}
  \centering
  \raggedright{\textbf{(c)}}
  \includegraphics[width=\linewidth]{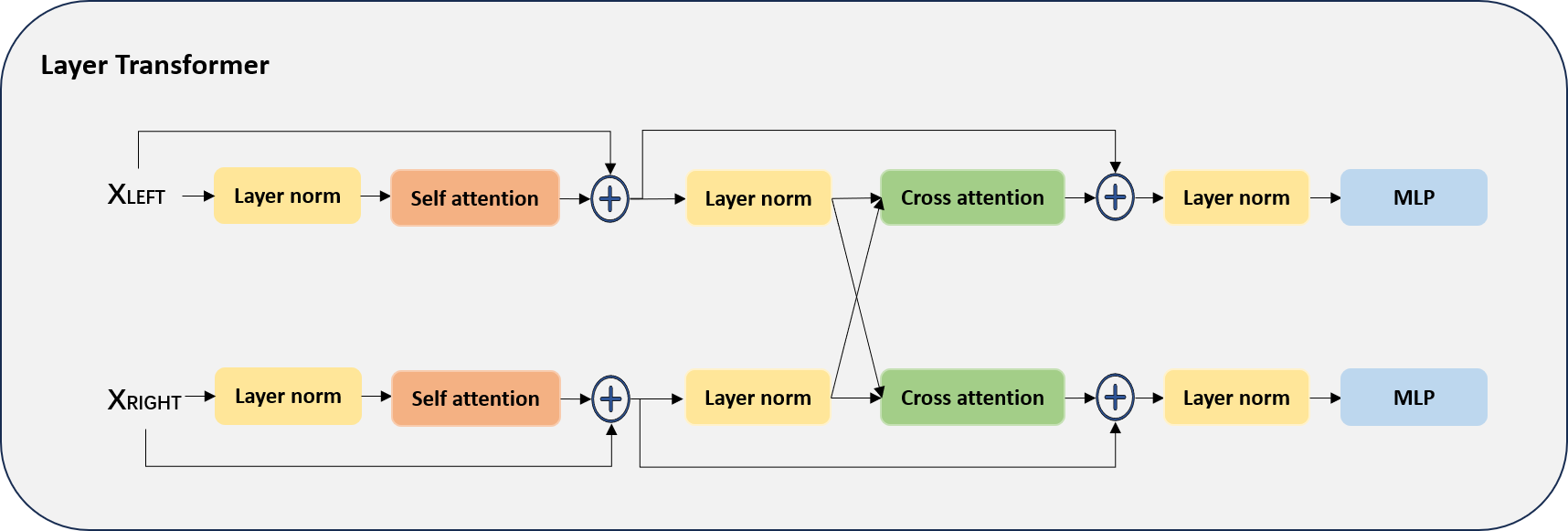}
\end{minipage}

\caption{\textbf{(a) Model architecture:} Blocks of identical color denote parameter sharing. The processing pipeline is independently applied to each channel of the binaural signal, producing a representation for each channel. These channel-specific representations are subsequently averaged and passed through a linear projection layer to estimate the intelligibility score. \textbf{(b) and (c)} illustrate the binaural processing blocks employed in the Temporal Transformer and the Layer Transformer.}
\label{fig:res}
\end{figure*}

\subsection{Model Architecture}
Fig. \ref{fig:res} illustrates the proposed \textbf{non-intrusive speech intelligibility prediction model architecture}, which incorporates a \textbf{Speech Enhancer} to generate a parallel enhanced signal pathway alongside the original noisy signal pathway. Leveraging recent advances in speech enhancement, the enhanced signals preserve informative speech characteristics. The enhanced pathway enriches model training with additional intelligibility information, addressing the common challenge of missing clean reference signals in real-world conditions. The binaural processing blocks within the temporal and layer transformers further demonstrate the use of cross-attention between corresponding channels of the noisy and enhanced signals.

\textbf{Feature Encoder Layer}: As shown in Fig. \ref{fig:res}(a), the noisy input is processed by the Speech Enhancer to generate an enhanced signal pathway parallel to the original noisy pathway. Both signals are then encoded by a pre-trained frozen SFM Encoder to extract features. To reduce computational complexity, the SFM features are average-pooled and downsampled by a factor of 20 before entering the Temporal Transformer, followed by a linear projection layer mapping the features into a 384-dimensional intermediate representation.

\textbf{Temporal Transformer}: The representations from both noisy and enhanced signals are processed by a Temporal Transformer to capture long-range temporal dependencies. As shown in Fig. \ref{fig:res}(b), each channel of the noisy and enhanced signals is first processed through a self-attention block. Cross-attention is then applied between corresponding channels to exploit supplementary intelligibility cues from the enhanced signal. To account for binaural contributions, a binaural cross-attention block is used \cite{cpc2model}. Finally, average pooling reduces the output to a tensor of size 1 × 384.

\textbf{Audiogram Projection Layer}: In parallel, the listener’s audiogram, measured at six frequencies (250 Hz, 500 Hz, 1000 Hz, 2000 Hz, 4000 Hz, and 6000 Hz), is passed through a linear projection layer to map it into a 384-dimensional space. The resulting representation is then concatenated with the output from the Temporal Transformer, producing a representation of size 2 × 384. 

\textbf{Layer Transformer}: As shown in Fig. \ref{fig:res}(c), a Layer Transformer block is then applied to process the concatenated tensor and extract intelligibility-related representations across layers. A global average pooling operation is performed along the layer axis, yielding a compact representation of size 1 × 384.

\textbf{Prediction Layer}: The processing pipeline is independently applied to each channel of the binaural signal, generating a channel-specific representation for each ear. These representations are subsequently averaged and passed through a linear projection layer followed by a sigmoid activation to produce the final intelligibility score. The score is scaled by a factor of 100 to match the range of the target intelligibility scores.

\subsection{Speech Enhancers}
In this work, we incorporate three SOTA DNN-based Speech Enhancers into our proposed architecture for non-intrusive intelligibility prediction:

\textbf{ZipEnhancer} \cite{ZipEnhancer} is a dual-path Zipformer-based monaural speech enhancement model that integrates time and frequency domain Down–up sampling to achieve effective noise suppression. We use the pretrained lightweight variant ZipEnhancer(S), which achieves a WB-PESQ of 3.69 on the DNS Challenge 2020 official test set without reverberation \cite{dns2020}. \footnote{\url{https://zipenhancer.github.io/ZipEnhancer}}

\textbf{MP-SENet} \cite{MP-SENET} is a codec-based speech enhancement model that jointly denoises magnitude and phase spectra using convolution-augmented transformers and parallel decoders, trained with multi-level spectral and waveform losses. We use the provided pretrained model, which achieves a WB-PESQ of 3.60 on the same DNS Challenge 2020 test set. \footnote{\url{https://github.com/yxlu-0102/MP-SENet}}

\textbf{FRCRN} \cite{FRCRN} is a single-channel noise reduction method which can be used for enhancing speech in various noise environments, which is developed based on a new framework of Convolutional Recurrent Encoder-Decoder (CRED). We use the pretrained FRCRN\_SE\_16K model, which processes both input and output of 16 kHz time-domain waveforms and achieves a WB-PESQ of 3.23 on the same DNS Challenge 2020 test set.\footnote{\url{https://github.com/alibabasglab/FRCRN}} 

For brevity, in the remainder of this paper we refer to the specific speech enhancers with their corresponding pretrained checkpoints as ZipEnhancer, MP-SENet, and FRCRN.

\renewcommand{\arraystretch}{1.05} 
\begin{table*}[thbp]
\centering
\caption{Intelligibility prediction performance (RMSE and NCC) on CPC3 development and evaluation sets. Baselines from Clarity Prediction Challenge are shown alongside our models with clean speech reference, different speech enhancers, and their ensembles. The ZipEnhancer + MP-SENet ensemble achieved the best performance.}
\begin{tabular}{lcccc}
\toprule
\textbf{Model} & \textbf{Dev RMSE} & \textbf{Dev NCC} & \textbf{Eval RMSE} & \textbf{Eval NCC} \\
\midrule
CPC3 Baseline (HASPI) & 28.00 & 0.72 & 29.47 & 0.70 \\
CPC2 Champion & 24.15 & 0.81 & 26.42 & 0.78 \\
\midrule
Ref-based Model & 23.86 & 0.82 & 26.18 & 0.79 \\
ZipEnhancer Model & 23.59 & 0.82 & 25.87 & 0.79 \\
MP-SENet Model & 23.63 & 0.82 & 26.14 & 0.79 \\
FRCRN Model & 24.15 & 0.80 & 26.58 & 0.78 \\
\midrule
ZipEnhancer + MP-SENet & \textbf{23.21} & \textbf{0.83} & \textbf{25.60} & \textbf{0.79} \\
ZipEnhancer + FRCRN & 23.53 & 0.82 & 25.90 & 0.79 \\
MP-SENet + FRCRN & 23.66 & 0.82 & 26.16 & 0.78 \\
ZipEnhancer + MP-SENet + FRCRN & 23.43 & 0.82 & 25.87 & 0.79 \\
\bottomrule
\end{tabular}
\label{tab: CPC3 intelligibility_results}
\end{table*}

\section{Experimental Setup}

\subsection{Dataset}
In our experiments, we employ two binaural speech intelligibility datasets: CPC3 Dataset and Arehart Dataset.

\textbf{CPC3 Dataset}: The CPC3 dataset \cite{cpc3_data} from the 3rd Clarity Prediction Challenge includes hearing aid (HA) output signals, clean speech references, intelligibility scores, and audiograms from 33 hearing-impaired (HI) listeners. It is split into 15,464 training, 926 development, and 7,674 evaluation samples, all sampled at 32 kHz. Both the development and evaluation sets were treated as test sets in our experiments.

\textbf{Arehart Dataset}: The Arehart dataset \cite{arehart_data} contains 8,100 binaural audio samples from 15 normal-hearing (NH) and 15 hearing-impaired (HI) listeners, each evaluated on 540 stimuli. Audio files are sampled at 22.05 kHz. For processing, all signals were resampled to 16 kHz. We split the HI data into 6,480 training samples (12 listeners) and 1,620 test samples (3 listeners) to prevent data leakage and improve generalization to unseen listeners.

For processing with the SFM and Speech Enhancer, all signals were resampled to 16 kHz to ensure consistency with the input requirements of both components.

\subsection{SFM with Encoder Layer Selection}
In this study, we select the state-of-the-art SFM, \textbf{Parakeet}, which combines a FastConformer encoder \cite{Fastconformer} with a Token-and-Duration Transducer (TDT) decoder \cite{TDT}. We use the parakeet-tdt-0.6b-v2 variant, updated by NVIDIA in May 2025, containing 600 million parameters and trained on 120,000 hours of English speech from the Granary dataset \cite{granary}. Based on prior research \cite{sfm_layer_selection}, we systematically evaluated all 23 encoder layers and found that layer 18 achieved the lowest RMSE for intelligibility prediction, which we use for all experiments in this study.

\renewcommand{\arraystretch}{1.05} 
\begin{table}[t]
\centering
\caption{Comparison of CPC2 Champion and our model (ZipEnhancer + MP-SENet) on the Arehart test set in terms of intelligibility prediction performance (RMSE and NCC). Best results are highlighted in bold}
\begin{tabular}{lcc}
\toprule
\textbf{Model} & \textbf{Test RMSE} & \textbf{Test NCC} \\
\midrule
CPC2 Champion              & 27.00 & 0.72 \\
ZipEnhancer + MP-SENet     & \textbf{26.12} & \textbf{0.73} \\
\bottomrule
\end{tabular}
\label{tab:arehart_intelligibility_results}
\end{table}

\renewcommand{\arraystretch}{1.05}
\begin{table*}[t]
\centering
\caption{Cross Dataset intelligibility prediction performance (RMSE and NCC) on CPC3 Eval and Arehart Test. Results are shown for CPC2 Champion and our best ensemble (ZipEnhancer + MP-SENet) under different training and augmentation settings. NH denotes adding 8100 normal hearing samples. 2-clips denotes augmentation of concatenating two random clips to increase 540 samples per listener.}
\begin{tabular}{llcccc}
\toprule
\multirow{2}{*}{\textbf{Model}} & \multirow{2}{*}{\textbf{Training Set}} & 
\multicolumn{2}{c}{\textbf{CPC3 Eval}} & 
\multicolumn{2}{c}{\textbf{Arehart Test}} \\
\cmidrule(lr){3-4} \cmidrule(lr){5-6}
 & & \textbf{RMSE} & \textbf{NCC} & \textbf{RMSE} & \textbf{NCC} \\
\midrule
\multirow{2}{*}{CPC2 Champion} 
  & CPC3                  & 26.42 & 0.78 & 32.86 & 0.62 \\
  & Arehart               & 27.87 & 0.73 & 27.00 & 0.72 \\
\midrule
\multirow{2}{*}{ZipEnhancer + MP-SENet} 
  & CPC3                  & 25.60 & 0.79 & 31.52 & 0.64 \\
  & Arehart               & 27.44 & 0.74 & 26.12 & 0.73 \\
\midrule
\multirow{4}{*}{ZipEnhancer + MP-SENet + Data Aug} 
  & CPC3 + NH         & 25.83 & 0.77 & 30.14 & 0.67 \\
  & Arehart + NH      & 28.22 & 0.73 & 26.93 & 0.71 \\
  & CPC3 + 2-clips    & \textbf{25.33} & \textbf{0.80} & 28.48 & 0.72 \\
  & Arehart + 2-clips & 26.81 & 0.77 & \textbf{25.43} & \textbf{0.78} \\
\bottomrule
\end{tabular}
\label{tab:cross_dataset_results}
\end{table*}

\subsection{Training, Evaluation and Baseline}
All experiments are trained with Huber loss for 50 epochs with a batch size of 128 using Adam optimizer with a learning rate of 4e-5, $\beta_1 = 0.9$ and $\beta_2 = 0.98$. For training, we perform a 3-fold train/validation split within the training set, randomly partitioning it into 80\% for training and 20\% for validation in each fold. For each fold, the model checkpoint with the lowest RMSE on the validation data is selected. During inference on the test set, the final intelligibility prediction is obtained by averaging the prediction scores from all 3 folds.

The evaluation metrics used in this study are the root mean square error (RMSE) and the normalized Pearson correlation coefficient (NCC). As baselines, we adopted the HASPI model (CPC3 baseline) [1] and the CPC2 champion model, the best-performing non-intrusive intelligibility prediction model to date.

\subsection{Data Augmentation}
To address the degraded generalization observed in preliminary cross-dataset evaluation, we employed data augmentation strategies to improve robustness. Two types of augmentation were implemented. First, we included samples from 15 NH listeners in the Arehart dataset during training, thereby exposing the model to a wider range of listener characteristics. Second, we randomly concatenated two audio clips from the same listener with a short silence period in between, generating 540 additional training samples per listener. For these concatenated samples, the corresponding intelligibility score was computed as the average of the two original scores.

\section{Results}

\subsection{CPC3 Dataset}
Table~\ref{tab: CPC3 intelligibility_results} summarizes the intelligibility prediction results (RMSE and NCC) on the CPC3 development and evaluation sets. Considering the single-enhancer models, the reference-based model, and the CPC2 Champion baseline, the overall performance ranking is: $ \text{ZipEnhancer} > \text{MP-SENet} > \text{Ref-based} > \text{CPC2 Champion} > \text{FRCRN} $. The reference-based (intrusive) model surpasses the CPC2 Champion baseline, underscoring the value of incorporating a clean reference pathway during training. Beyond this, models with proper selection of speech enhancers can further exceed the the performance of ref-based model. Among single-enhancer models, ZipEnhancer achieves the best results, followed by MP-SENet, while FRCRN performs the worst, falling below the CPC2 Champion baseline. The ensemble of ZipEnhancer and MP-SENet yields the strongest overall performance with improvements in RMSE of 0.94 and 0.82, on the development and evaluation set respectively, over the CPC2 Champion baseline. For all subsequent experiments, this ensemble configuration is consistently used.

\subsection{Arehart Dataset}

Unlike CPC3 dataset, the Arehart dataset does not include reference signals, limiting training and evaluation to non-intrusive intelligibility prediction. As shown in Table~\ref{tab:arehart_intelligibility_results}, the model with ZipEnhancer + MP-SENet ensemble achieves improvements of 0.88 RMSE over the CPC2 Champion baseline, consistent with the performance gains observed on CPC3 dataset.

\subsection{Cross Dataset Generalization}
Regarding cross-dataset generalization (Table~\ref{tab:cross_dataset_results}), both the CPC2 Champion and our ZipEnhancer + MP-SENet model exhibit performance degradation on unseen datasets. Augmenting with NH data yields minimal benefit, indicating that incorporating normal-hearing samples during training does not improve performance. In contrast, 2-clips augmentation, which enriches HI data per listener, consistently improves performance and robustness on unseen data underscoring the value of proper data augmentation for cross-dataset generalization.

\section{Discussion}

Reference-based (intrusive) methods use clean speech to help models learn richer mappings from noisy to clean features, thereby enhancing prediction performance, but their reliance on unavailable reference signals limits real-world applicability. Our model architecture, by leveraging strong speech enhancers, enables reliable non-intrusive intelligibility prediction in real-world scenarios without reference signals. Among single-enhancer models, ZipEnhancer performs the best, followed by MP-SENet, while FRCRN performs the worst, even below the CPC2 Champion baseline. This emphasizes that the choice of speech enhancer is critical in our framework, with prediction performance positively correlating with each enhancer’s speech enhancement quality (WB-PESQ: $ \text{ZipEnhancer 3.69} > \text{MP-SENet 3.60} >> \text{FRCRN 3.23} $). The ZipEnhancer + MP-SENet ensemble achieves superior intelligibility prediction by providing the prediction model with richer and more reliable cues than a single enhancer for estimating intelligibility.

With the evolution from CPC1 \cite{cpc1} and CPC2 \cite{cpc2} to CPC3, the test datasets now encompass greater variability in listeners and systems, heightening the importance of achieving robust generalization to unseen listeners and setting \cite{2025_nonintrusive_trends}. Poor cross-dataset generalization in speech intelligibility prediction stems from differences in listener profiles, recording conditions, and speech materials between datasets like CPC3 and Arehart. Models trained on a single dataset tend to capture dataset-specific characteristics, resulting in reduced performance when applied to a different dataset. Our experiment indicate that adding NH data provides minimal benefit, as it does not reflect HI listener variability. In contrast, 2-clips augmentation, by combining two random clips from the same HI listener, enriches speech diversity and acoustic contexts, enabling the model to learn more generalizable intelligibility cues and improving performance and robustness on unseen datasets.


\section{Conclusion}
We proposed a non-intrusive intelligibility prediction framework that leverages strong speech enhancers to form a parallel enhanced-signal pathway, enabling the model to extract additional intelligibility cues beyond the noisy input. This approach consistently outperforms the CPC2 Champion baseline across diverse datasets. While intrusive models benefit from reference signals, our results show that carefully chosen and ensembled enhancers provide a robust non-intrusive alternative for real-world scenarios where references are unavailable. In addition, a simple 2-clips augmentation mitigates cross-dataset generalization issues by improving transfer to unseen listeners and conditions. Together, these findings highlight the potential of enhancer-guided non-intrusive architectures to advance practical intelligibility prediction and support future hearing aid technologies.

\bibliographystyle{IEEEtran}
\bibliography{strings,refs}

\end{document}